\documentclass[aps,prl,groupedaddress,twocolumn,superscriptaddress,showkeys,showpacs]{revtex4-1}
\usepackage{amsfonts}
\usepackage{graphicx}
\usepackage{mathcomp}
\usepackage{wasysym}

\begin{document}

\title{Ultra-Broadband Microwave Frequency-Comb Generation in Superconducting Resonators}

\author{R. P. Erickson}
\affiliation{National Institute of Standards and Technology, Boulder, Colorado 80305, USA}

\author{M. R. Vissers}
\affiliation{National Institute of Standards and Technology, Boulder, Colorado 80305, USA}

\author{M. Sandberg}
\affiliation{National Institute of Standards and Technology, Boulder, Colorado 80305, USA}

\author{S. R. Jefferts}
\affiliation{National Institute of Standards and Technology, Boulder, Colorado 80305, USA}

\author{D. P. Pappas}
\email[]{Electronic address: David.Pappas@NIST.gov}
\affiliation{National Institute of Standards and Technology, Boulder, Colorado 80305, USA}

\begin{abstract}
We have generated frequency combs spanning 0.5 to 20 GHz in superconducting $\lambda /2$-resonators at $T=3$ K. Thin films of niobium-titanium nitride enabled this development due to their low loss, high nonlinearity, low frequency-dispersion, and high critical temperature. The combs nucleate as sidebands around multiples of the pump frequency. Selection rules for the allowed frequency emission are calculated using perturbation theory and the measured spectrum is shown to agree with the theory. Sideband spacing is measured to be accurate to 1 part in $10^8$. The sidebands coalesce into a continuous comb structure observed to cover at least 6 octaves in frequency.
\end{abstract}


\maketitle

Frequency combs in the optical regime have become extremely useful in a wide range of applications including spectroscopy and frequency metrology \cite{CoddingtonPRA2010dualcomb,YuJKPS2013clock,HatiIEEE2013stateofart}. Recently, it was found that a strongly pumped, high-Q optical microcavity made from a nonlinear medium generates sidebands due to a combination of degenerate and non-degenerate four-wave mixing (FWM) 
\cite{DelHayeNature2007comb,DelhayPRL2011octavespanning,FosterOpticsExpress2011SiBasedComb} that cascade into a broadband frequency comb of photon energies in the regime of hundreds of THz. The peaks of these combs are extremely narrow and appear at frequencies dictated by selection rules for photon energy and momentum conservation. These devices are attractive because they have very narrow linewidths, are relatively simple, highly stable and controllable \cite{FangAPL2013combLineWidth,PappPRX2013control,DelhayePRL2008stabilization}, and can be divided down into the GHz range to achieve very accurate frequency references. However, the microcavities, typically consisting of toroidal silica structures \cite{KippenbergPhD2008thesis}, need to be pumped with high laser powers because of their intrinsically weak $\chi(3)$ optical Kerr nonlinearity. In addition, output over much more than a single octave in frequency is difficult to obtain from these structures due to frequency dispersion from material and geometric factors, which make the modes non-equidistant.

These Kerr combs continue to be the focus of extensive theoretical analysis to understand the nonlinear dynamics that give rise to their threshold of stability, mechanism of cascade, amplitude of responsiveness, and maximum spectral bandwidth \cite{AghaOpticsExpress2009theoretical,ChemboPRA2010modal,ChemboPRL2010spectrum,
HanssonArXiv2013dynamics,GodeyArXiv2013stability}. Generation of these combs directly in the 1-20 GHz range would further simplify the instrumentation and potentially elucidate the dynamics involved by making them more accessible to direct measurement.

In the current work we transfer the nonlinear pumped cavity concept to the microwave regime in superconducting resonators and demonstrate broadband frequency comb generation over multiple octaves. This is achieved using niobium-titanium nitride (NbTiN) thin films and exploiting (i) the high quality factor $Q>10^7$ for a strong drive \cite{DayNature2003,*MazinAPL2006,BarendsAPL2010}, (ii) the large nonlinear kinetic inductance, and (iii) the lack of frequency dispersion \cite{HandbookOfSuperconductingMaterials2003}. The kinetic inductance, ${L_K}(t)={L_0}\{ 1 + \left[ I(t) \left/ I_* \right. \right] ^2 \}$, where $L_0$ is the geometric inductance and $I_*$ a normalization constant comparable to the critical current, arises from the stored kinetic energy of charge carriers.

\begin{figure}
\includegraphics[width=240pt, height=150pt]{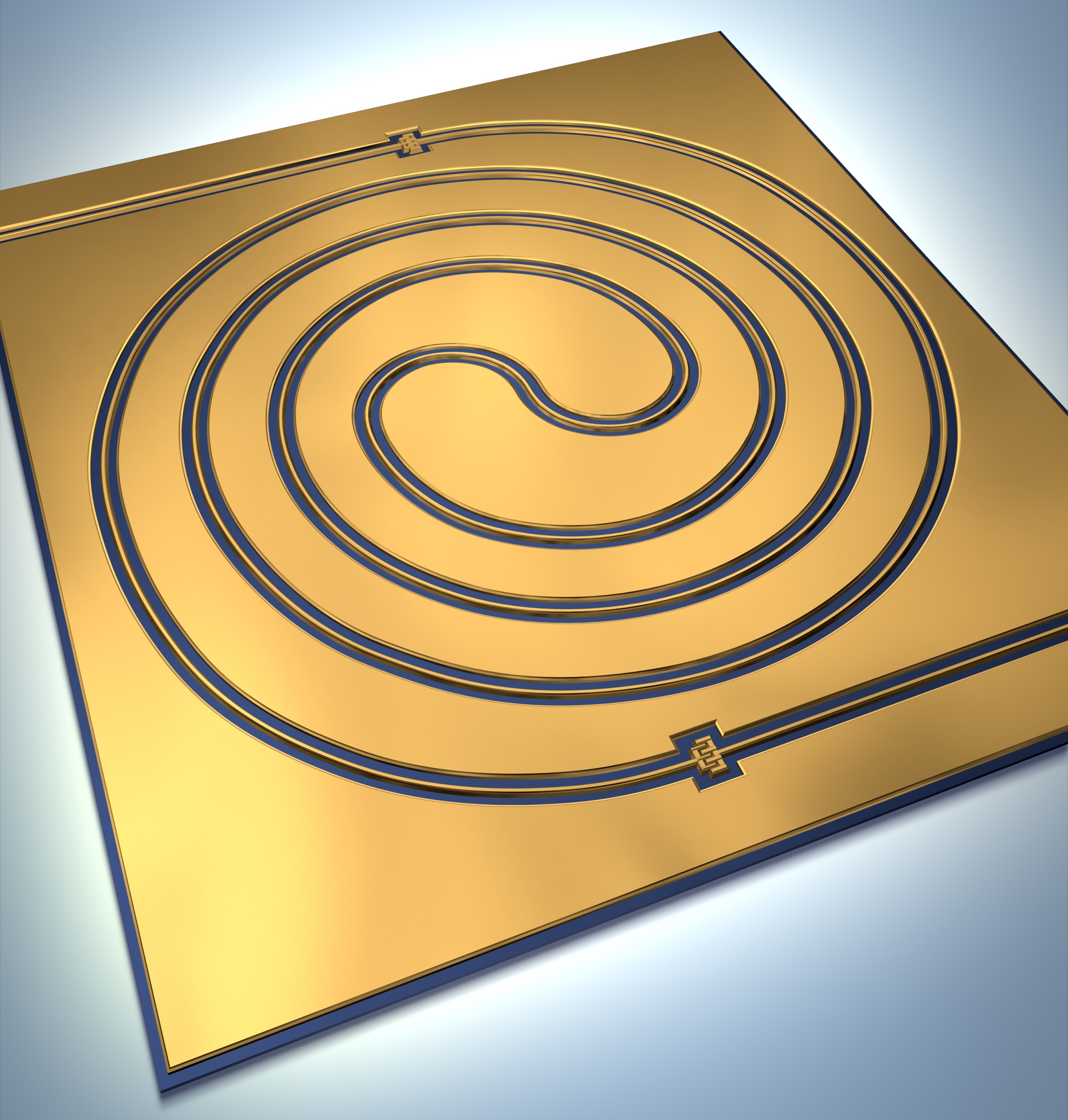}
\caption{\label{fig1}Artist's rendition (not to scale) of the superconducting frequency comb chip. The 25 cm long, $\lambda /2$ resonator is made in a coplanar waveguide (CPW) geometry with input and output ports on the top left and bottom right. The CPW has a $2 {\mu}m$ wide center strip and $2 {\mu}m$ wide gap. It is coupled to the ports by inter-digitated capacitors. The device is made from NbTiN (Au color) on a 2 cm x 2 cm intrinsic Si ($ > 20 k\Omega $) substrate.}
\end{figure}

For an excitation in the resonator, $I_{res}(t) = I\cos \omega t$, these resonators display a $\chi(3)$ Kerr-like behavior that generates odd harmonics of the excitation because the voltage drop across the resonator inductance, ${L_K}(t) dI(t) \left/ dt \right. $, leads to initial nonlinear response terms such as ${I_{res}}{(t)^3} \propto 3\cos \omega t + \cos 3\omega t$. More precisely, modeled as a simple transmission line, the current of the resonator satisfies a nonlinear, second-order differential equation that may be expressed as
\begin{equation}
\frac{{d}^2 {A}(t)}{{d t}^2} + \omega _0^{2} A(t) + \frac{1}{3} \frac{{d}^2 {A(t)}^3}{{d t}^2} = F\cos {\omega _P}t , \label{diffEq}
\end{equation}
where ${A}(t)=I_{res}(t)\left/ I_* \right. $ is a dimensionless amplitude, $\omega _0$ is the resonator fundamental frequency, and $F$ is the effective driving force of an external pump of frequency $\omega _P$. A derivation of Eq. (\ref{diffEq}) and an analysis of characteristics of its broadband-comb solution are given in an online supplement \footnote{See Supplemental Material at URL for design, fabrication, theoretical analysis, and additional video material.}.

Geometrically engineered frequency dispersion has been used to make wide-band traveling-wave amplifiers in coplanar waveguide (CPW) transmission lines of NbTiN \cite{EomNature2012amplifier}. Such engineering can be applied to these materials because there is no intrinsic frequency dispersion (within 5\% measurement accuracy) up to $f_{\max } \cong 2{ \Delta \left/ h \right. } \times 66\% $ \cite{Javadi1992}. For NbTiN this corresponds to frequencies on the order of 600 GHz.  This lack of dispersion creates the possibility of generating frequency combs with multiple octaves of bandwidth, which will provide a powerful tool in the rapidly growing field of superconducting electronics.

Half-wave CPW resonators fabricated from 20 nm thick NbTiN films were used, and comb generation was observed up to $T \sim 6$ K due to the high $T_C \sim 13$ K of the films. The geometries used included both transmission, illustrated in Fig. \ref{fig1} and modeled via Eq. (\ref{diffEq}), and reflection, described in \cite{Wisbey2010}. The unperturbed fundamental resonant frequency for these resonators is given by $f_0 = \omega _0 \left/ 2\pi \right. = c \sqrt{1 - \alpha } \left/ (2 l n_{eff}) \right. $, where $l$ is the length, $n_{eff} = 2.6$ is the effective index of refraction for a CPW on Si, and $\alpha = 0.93 $ is the kinetic inductance fraction as determined from the frequency shift of a test resonator. We were thus able to set $f_0$ in the range of 15 MHz up to 6 GHz with easily achievable lengths from 1 m down to 2.5 cm. For clarity and brevity, the discussion here is restricted to a single device, a 25 cm long NbTiN resonator with $f_0 = 59.738181(1)$ MHz. The design, fabrication, and theoretical analysis are described in the online supplement. 

Frequency comb emission is excited in these devices by applying a pump tone at frequency  $f_P = \omega _P \left/ 2\pi \right.$. For $f_P = N f_0 + \delta f$, power is coupled into the resonator at both $f_P$ and $f_0$ when the detuning, $\delta f$, is small. Moreover, in addition to output at the fundamental and pump frequencies, the nonlinear resonance is characterized by a new set of subharmonic states, distinct from natural modes of the resonator cavity, that form at odd harmonics of the pump frequency. Unlike the case of linear response, proximity of $f_P$ to $f_0$ is not a requirement. Quite the contrary, $f_P$ can be spectrally distant from $f_0$, with strong nonlinear response elicited as $\delta f$ is decreased. 

The initial current induced in the resonator is \begin{equation}
{I_{res}}(t) = I_0 \cos 2\pi f_0 t + I_P \cos 2\pi f_P t , \label{currentEq}
\end{equation}
where $I_0$ and $I_P$ depend on the detuning, pump power, nonlinearity, and strength of the coupling. In particular, for constant pump power, as $f_P$, and hence $\delta f$, is decreased the current induced in the resonator renormalizes $f_0$ downward due to the dependence of the kinetic inductance on current \cite{Karpov2004}. As $\delta f$ approaches zero, state bifurcation invariably occurs and the resonator jumps back to a quiescent state at some critical frequency $f_P = f_{crit}$ \cite{Swenson2013}. However, prior to this event enough power may be coupled into the resonator to cross the parametric oscillation threshold wherein the gain exceeds cavity losses. This condition permits steady-state generation of a full range of frequency sidebands and FWM products \cite{Swenson2013}, seeded by the principal subharmonic states.

\begin{figure}
\includegraphics[width=220pt, height=130pt]{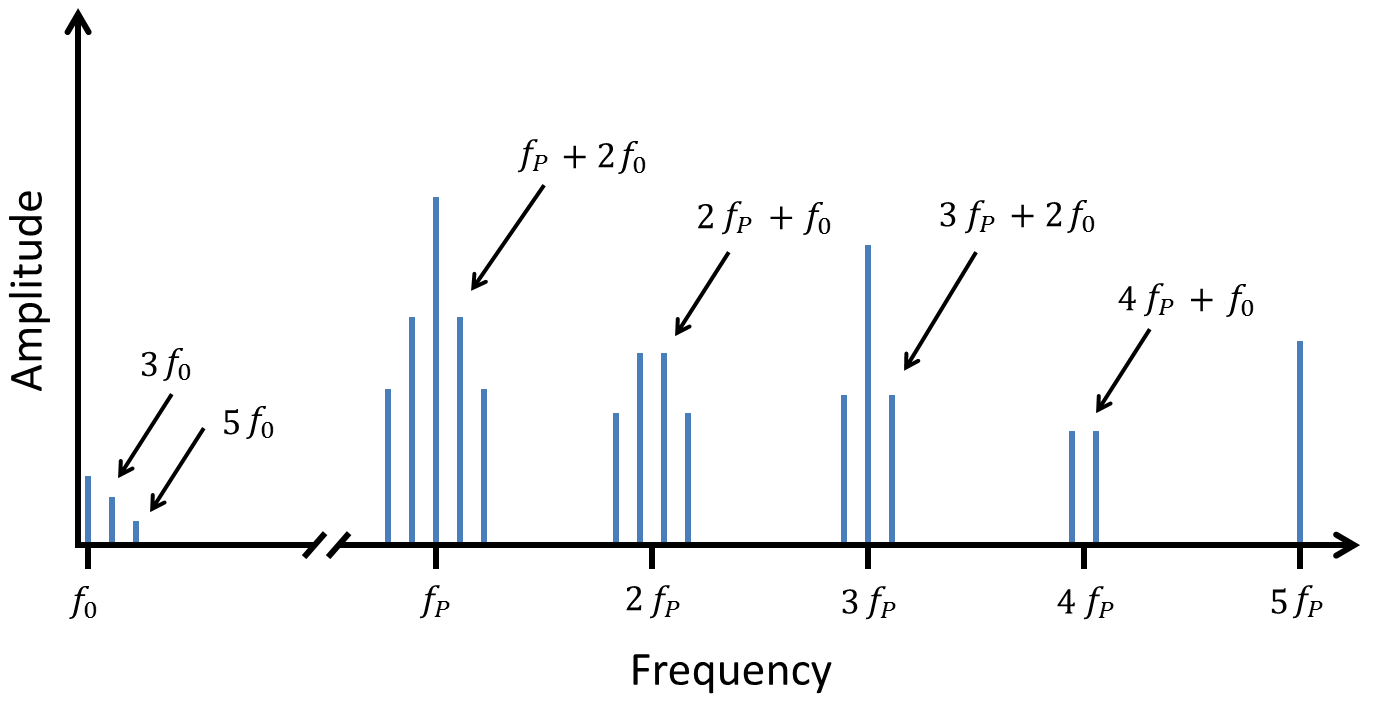}
\caption{\label{fig2}Emission spectrum predicted by perturbation theory for the multi-octave frequency comb, with fundamental frequency $f_0$ and the pump at $f_P$. Illustrated are the frequencies from Eq. (\ref{currentEq}), associated sidebands from Eqs. (\ref{allowedFrequencies}), and extra peaks of cascade spaced at $2 f_0$.}
\end{figure}

The selection rules for allowed response frequencies and their spacing arise due to mixing between the pump and fundamental frequencies, as when Eq. (\ref{currentEq}) is cubed and the pertinent trigonometric identity is applied. This mixing is explicitly derived in the online supplement, where second-order perturbation theory is applied to Eq. (\ref{diffEq}) to show that 
\begin{enumerate}
	\item[(i)]Odd harmonics of the pump, $M w_P$, where $M = 1, 3, 5, \ldots $, are permitted as principal teeth of the comb,
	\item[(ii)]Sideband teeth spaced at $2 f_0$ are generated around the odd pump harmonics,
	\item[(iii)]Even harmonics of the pump, with $M = 0, 2, 4, \ldots $ are forbidden, and
	\item[(iv)]Sideband teeth spaced at $2 f_0$ are generated around the absent even pump harmonics.
\end{enumerate}
For integer $M\ge 0$, the allowed frequencies of these rules are summarized as 
\begin{eqnarray}
f_0, 3 f_0, 5 f_0, \ldots ; M = 0 \nonumber \\
M f_P, M f_P \pm 2 f_0, M f_P \pm 4 f_0, \ldots ; M odd \nonumber \\
M f_P \pm f_0, M f_P \pm 3 f_0, \ldots ; M even \label{allowedFrequencies}
\end{eqnarray}
The expected emission spectrum from Eqs. (\ref{allowedFrequencies}) is illustrated in the sketch of Fig. \ref{fig2}.

\begin{figure}
\includegraphics[width=240pt, height=150pt]{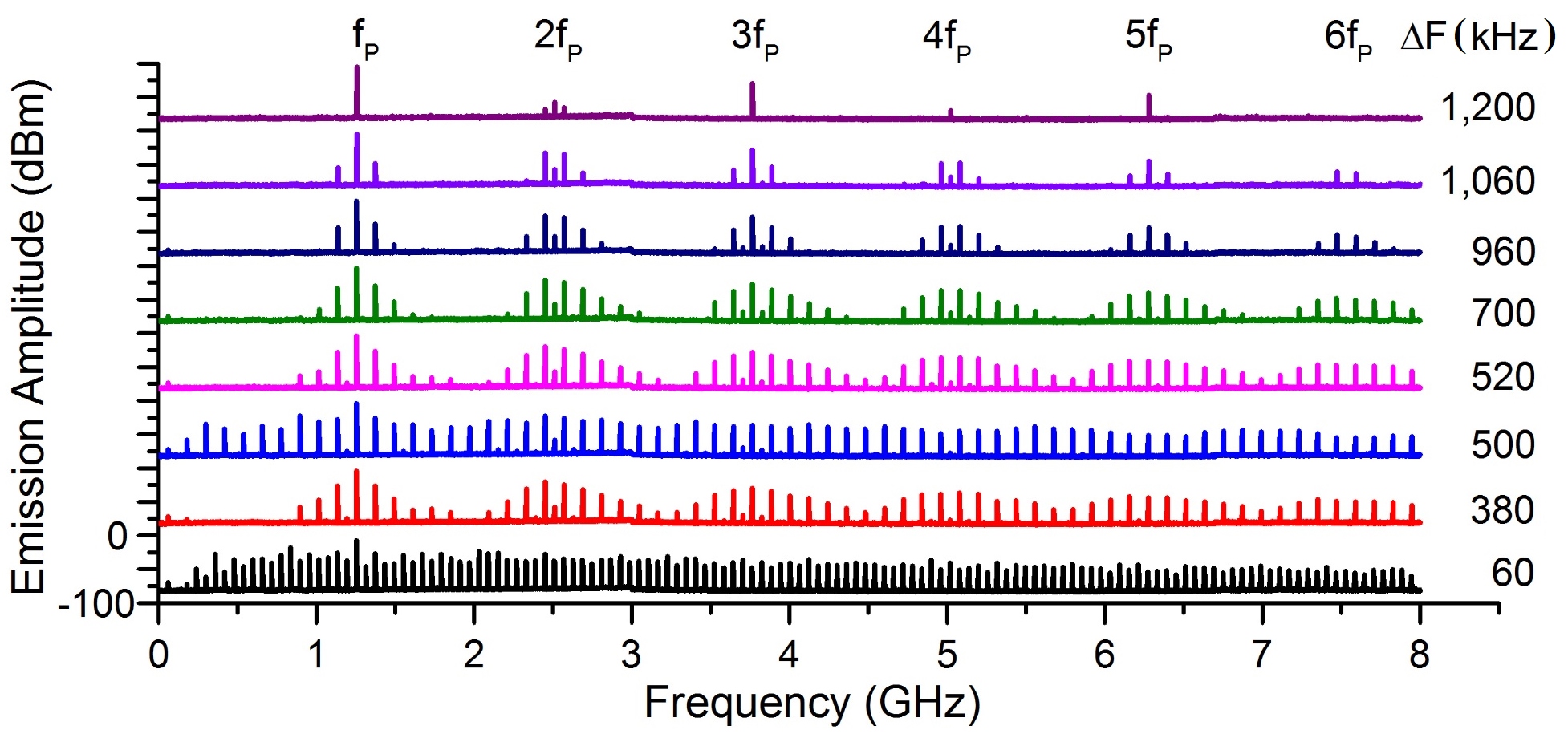}
\caption{\label{fig3}Evolution of the emission spectrum as the difference, $\Delta F = f_P - f_{crit}$, between the pump frequency $f_P$ and the critical bifurcation frequency, $f_{crit}=1254.7$  MHz, is decreased. The pump is close to the $N=21$ multiple of the resonator fundamental, $f_0$. The sample is held at $T=3$ K, with pump power of the feedline held constant at -28 dBm. Traces are offset vertically for clarity. The signal has been amplified by 30 dB. (Finer detail as $\Delta F \rightarrow 0$ may be found in the successive frames of the online video, where each frame is equivalent to a trace of the above stack plot.)}
\end{figure}

Measurements of the frequency emission spectrum were conducted at low temperature, $T=3$ K, in a magnetically unshielded copper box. An RF signal generator was connected to the input to excite the system and a spectrum analyzer was connected to the output. The experiments described here used pump powers of -28(1) dBm with detuning $\delta f \sim 100$ kHz.

Figure \ref{fig3} shows a typical evolution of the spectrum as $f_p$ is decreased to the point of bifurcation. For convenience, we define $\Delta f = f_P - f_{crit}$. Far from bifurcation, above $\Delta F = 1200$ kHz, we see predominantly odd multiples of the pump in the spectrum, i.e., just the principal teeth of odd sidebands. Some emission at $2 f_P$ and $4 f_p$ is observed, albeit at -25 dB relative to the odd harmonics, and can be accounted for by distortion in the amplifier and/or parasitic slot-line modes. The amplifiers also had a low frequency cutoff below 500 MHz, thereby filtering out the response at $f_0 \cong 60$ MHz.

As $\Delta F$ approaches 1200 kHz, sideband teeth spaced by $2 f_0$ are first observed around the $2 f_P$ location, shown in the top spectrum of Fig. \ref{fig3}. At $\Delta F = 1,060$ kHz sideband teeth begin to appear around both even and odd multiples of the pump, as described by Eq. (\ref{allowedFrequencies}). The full width at half-maximum (FWHM) of the sideband peaks is the same across the spectrum and measured to be 1.1(0.1) Hz, limited most likely by the resolution bandwidth of the spectrum analyzer. This is nearly an order of magnitude less than that expected from a $Q=10^7$ resonator, consistent with states that do not couple to a dissipative reservoir. These sidebands continue to develop down to $\Delta F = 520$ kHz. As $\Delta F$ continues to decrease and the system is pushed closer to bifurcation, the sideband structure undergoes a sudden transition, coalescing into a continuous, broadband comb structure at $\Delta F = 500$ kHz. This structure can persist to well above 20 GHz, depending on the specific $f_P$ and power used. The upper limit of the response readout is currently limited by the connectors (SMA) used on the system, but even with this configuration, we see cascades spanning at least 6 octaves in frequency.

The system undergoes two more transitions as it nears perfect tuning. The second transition, at $\Delta F = 380$ kHz, occurs where it switches back into a modulated broadband comb, and the final, third transition at $\Delta F = 60$ kHz sees it coalesce again into a smooth spectrum with a modified spacing between sidebands of $1 \times f_0$. The comb then collapses as $\Delta F \rightarrow 0$ and the system goes past a bifurcation point.

The change in the sideband spacing is consistent with period doubling that typically occurs in nonlinear systems as they go through bifurcations \cite{JordanAndSmith2007,*Strogatz2000}. This interpretation is supported by data taken for varying values of pump power, where period doubling is always observed just before bifurcation, even at low power. This observation rules out other effects such as amplifier saturation or power-dependent modes in the CPW. This evolution of behavior is repeated, albeit in a slightly modified nature, for pumps with different subharmonic matching. We also note that for phases with continuous, coalesced sidebands (around $\Delta F = 500$ and 60 kHz), multiple satellite peaks appear around the comb teeth with frequency spacing $\sim \delta f$. This indicates that separate sidebands from the various multiples of the pump are beating together, owing to the many-octave extent of the entire broadband structure. 

\begin{figure}
\includegraphics[width=240pt, height=300pt]{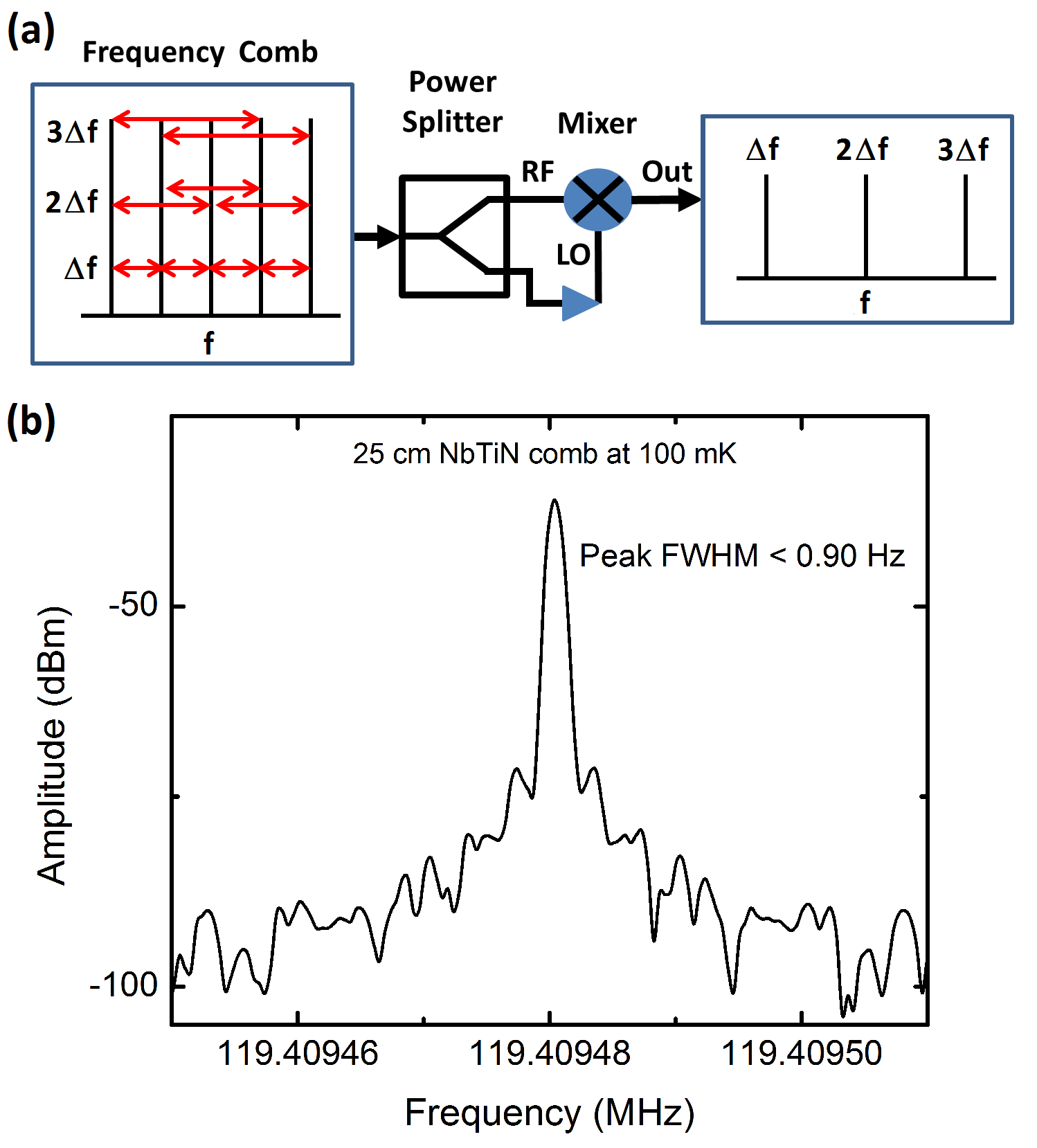}
\caption{\label{fig4}Mixing measurement of the sideband peak spacing. Schematic is shown in panel (a), and  output with the comb pumped at $\Delta F =520$ kHz is panel (b). The device was measured at 100 mK in an adiabatic demagnetization refrigerator for temperature stability.}
\end{figure}

In order to accurately measure the spacing of the sidebands, a nonlinear mixing process is employed. The scheme is shown in Fig. \ref{fig4}. In this measurement, the output signal is split into two components. One component is amplified and applied to the local oscillator (LO) input of a wide-band mixer. The other component is then applied to the RF input, where each tooth of the comb is compared to the inputs. Each comb tooth therefore acts as a reference for all other teeth, giving an output that reflects the overall comb periodicity. The appearance of peaks in this spectrum also shows that the different sidebands are phase coherent.

In the mixing measurement, we find a dominant, single-valued peak at $2 f_0$ for pump subharmonics down to $\Delta F = 520$ kHz and again at $\Delta F = 380$ kHz. Close to $\Delta F = 500$ kHz and $\Delta F = 60$ kHz we observe multiple beating from the extra satellites, with the dominant peak in the mixing data switching to $1\times f_0$ before the comb collapses. The FWHM of the main peak here is measured to be less than 1 Hz, corresponding to frequency resolution better than one part in $10^8$. Drift of the position of the peak on the order of 10 Hz occurs on the 10 to 100 second time scale, with characteristic jumps consistent with flux trapping in the CPW gap. The spectrum also shows mirroring around the main peak, consistent with coherent beating between the various sidebands. These results agree well with predictions from perturbation theory and the observed spectral response.

At present, we have modeled this system to second order in perturbation theory to understand the selection rules that define the spacing between sidebands. This analysis is outlined fully in the online supplement. However, the frequency cascade and existence of transitions that redefine the sideband spacing close to bifurcation are consistent with a highly correlated, nonlinear system that requires a full FWM and detailed balance calculation. We are currently exploring suitable models. The simplicity of these devices, their low dispersion, high nonlinearity, and the fact that they can be easily measured with standard RF techniques make them an exciting platform to study nonlinear phenomena.

In conclusion, we have demonstrated and theoretically modeled broadband frequency-comb generation in highly nonlinear superconducting resonant cavities. We have fabricated and tested multiple devices with different materials and free-spectral ranges and find highly reproducible and reliable behavior. The stability of the comb generation is expected to improve as magnetic shielding and multiple-octave feedback is added \cite{DelhayePRL2008stabilization}. The low loss and lack of dispersion allow for multiple decades of comb generation. Since the temperatures needed are achievable with a closed cycle He compressor, we expect that these devices will allow for relatively low-cost, frequency-agile devices in the near future.

\begin{acknowledgments}
For important insights, the authors are grateful for helpful discussions with Jiansong Gao, Pascal Del’Haye, Scott Diddams, Dave Howe, Dylan Williams, and Dan Slichter. This work was supported by DARPA and the NIST Quantum Information initiative. RPE acknowledges grant 60NANB14D024 from the US Dept. of Commerce, NIST. This work is property of the US Government and not subject to copyright.
\end{acknowledgments}

\bibliography{manuscript}

\end{document}